# SCALING LAWS IN PARTICLE PHYSICS AND ASTROPHYSICS

RUDOLF MURADYAN

*Dedicated to the Golden Jubilee (1961-2011) of publication of the article by Geoffrey Chew and Steven Frautschi in Phys. Rev. Lett. 7, 394, 1961, where a celebrated scaling law $J \approx m^2$ has been conjectured for spin/mass dependence of hadrons.*

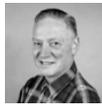 G. Chew    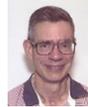 S. Frautschi

1. INTRODUCTION: WHAT IS SCALING?

Any polynomial *power law* $f(x) = c\, x^n$, where constant $c$ has a dimension

$$\dim c = \frac{\dim f}{(\dim x)^n}$$

exhibits the property of *scaling* or *scale invariance*. Usually $n$ is called scaling exponent. The word *scaling* express the fact that function $f$ is shape-invariant with respect to the *dilatation* transformation $x \to \lambda x$

$$f(\lambda x) = c(\lambda x)^n = \lambda^n f(x)$$

and this transformation preserves the shape of function $f$. We say, following Leonhard Euler, that $f$ is *homogeneous* of degree "$n$" if for any value of parameter $\lambda$ $f(\lambda x) = \lambda^n f(x)$. Differentiating this relation with respect to $\lambda$ and putting $\lambda = 1$ we obtain simple differential equation

$$x f'(x) = n f(x)$$

solution of which brings back to the polynomial power scaling law.

There are tremendously many different scaling laws in Nature. The most important of them can be revealed by Google search of scaling site:nobelprize.org in the official site of Nobel Foundation, where nearly 100 results appears. Ten of them are shown below:

1. Jerome I. Friedman - Nobel Lecture
2. Daniel C. Tsui - Nobel Lecture

3. Gerardus 't Hooft - Nobel Lecture

4. Henry W. Kendall - Nobel Lecture

5. Pierre-Gilles de Gennes - Nobel Lecture

6. Jack Steinberger - Nobel Lecture

7. Richard E. Taylor - Nobel Lecture

8. Frank Wilczek - Nobel Lecture

9. Kenneth G. Wilson - Nobel Lecture

10. David J. Gross - Autobiography

Disclosure of scaling relationship between observable quantities gives direct information about dynamics of natural phenomenon. This is the main reason why *scaling* plays a key role in the methodology of natural sciences.

In this talk, we will consider several diverse scaling laws in particle physics and astrophysics. The Part I is based mainly on the work [1]. Part II is dedicated to the to the extension of Chew-Frautschi hadronic spin/mass scaling relation to the realm of astronomical objects.

2. PART I. SCALING LAWS IN PARTICLE PHYSICS

Until to the first half of fifties of 20$^{th}$ century all known elementary particles were assumed structureless, i.e. point-like. Principal results about compositeness of nucleon were obtained in 1957 from elastic electron-proton scattering measurements at SLAC
(Stanford Linear Accelerator Center) and the work is still in progress. R. Hofstadter received the Nobel Prize in 1961 for the first results about charge and magnetization distributions in nucleons. These basic experiments clearly showed that protons and neutrons are non-elementary particles with proper electric and magnetic form factors.

In 1973 V. Matveev, R. M., and A. Tavkhelidze reported that asymptotic behavior of electromagnetic form factors of hadrons contains information about distribution and dynamics of *quarks* in hadrons and proposed dimensional Quark Counting Rules [1]. Concurrently dimensional QCR has been suggested and developed by Stanley Brodsky and his collaborators from SLAC [2].

Quark model was independently proposed by G. Zweig and M. Gell-Mann in 1964 on the basis of SU(3)-symmetry and generalizations of Sakata model. Sakata model is precursor of quark model and historically the SU(3)-symmetry first was introduced in Sakata model in 1959. Shoichi Sakata and his Kioyto theory group (M. Ikeda, S. Ogawa, Y. Ohnuki, Y. Yamaguchi,...) have been considered as *Marxists* and marginalized by mainstream American physicists ( as noted by David Kaiser, historian of science [3]).

The SLAC deep-inelastic electron-proton scattering experiments $e + p \rightarrow$ *anything* (1969) and its theoretical Interpretation by J. Bjorken, R. Feynman and others (V. Matveev, R. M., A. Tavkhelidze) had a considerable impact on understanding of structure of proton and relevant Bjorken scaling law. Bjorken's scaling law demonstrates that proton is made out of point-like objects of spin 1/2 (partons?).

Until development of dimensional Quark Counting Rules (1973) the total opinion has been prevailed that quarks are auxiliary, *mathematical devices* for description of SU(3)-symmetry. M. Gell-Mann and G. Zweig, creators of quark model, initially have been proponents of this point of view. Quark Counting Rules played significant role in clarifying that quarks are real, not plain "mathematical entities"! *)

In [1] a simple asymptotic behavior for hadronic electromagnetic form factors was suggested. For composite object *a* with $n_a$ constituents the corresponding form factor asymptotically must behaves as

$$F_a \approx \frac{1}{t^{n_a-1}}$$

where *t* is corresponding Mandelstam variable.

This relation shows that the more constituents has an object the faster is fall-off of the form factor.

---

*) As noted by A. Abbas [4] : "...both Gell-Mann and Zweig did not posit any physical reality onto these quarks. As per them – these quarks were just 'mathematical entities' and SU(3) was convenient mathematical 'trick' cooked up to do the job... During the Nobel Prize Ceremony Gell-Mann did present a lecture entitled 'Symmetry and currents in particle physics' on Dec 11, 1969 at Stockholm. All Nobel Laureates are supposed to present a written version of these lectures to be published. But for some inexplicable reason Gell-Mann Nobel Lecture was not written up for publication in the collection of lectures (Nobel Lecture (1963-1970)). However it is reputed that he had actually referred to quarks as mere 'mathematical entities' in his lecture."

Thereafter for pion ($n_\pi=2$), nucleon ($a_N=3$), deuteron ($n_d=6$), $He^3$ ($n_{He3}=9$) and $He^4$($n_{He4}=12$) form factors asymptotically behave as

$$F_\pi \approx \frac{1}{t}, \ F_N \approx \frac{1}{t^2}, \ F_d \approx \frac{1}{t^5}, \ F_{He^3} \approx \frac{1}{t^8}, \ F_{He^4} \approx \frac{1}{t^{11}}$$

and symbolically are depicted on Fig.1

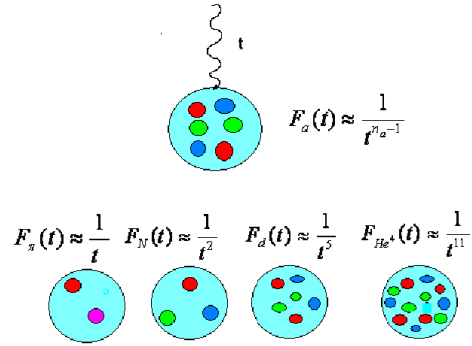

Fig.1 Asymptotic behavior of electromagnetic form factors of some few-quark composite systems according to the dimensional Quark Counting Rule .

For a two-body exclusive reaction *a+b -> c+d* Quark Counting Rules can be summarized as

$$\frac{d\sigma^{ab \to cd}(s,t)}{dt} \approx \frac{1}{s^{n-2}} f(\frac{t}{s})$$

where $n = n_a + n_b + n_c + n_d$ is the total number of quarks involved in the initial and final states of reaction, *s* and *t* are Mandelstam variables, *s* is the square of the total energy in the center-of-mass system, and *t* is the momentum transfer squared in the s channel; *f* is dimensionless function, depending on the details of the dynamics of process.

Some characteristic processes are depicted on Fig.2-Fig.3

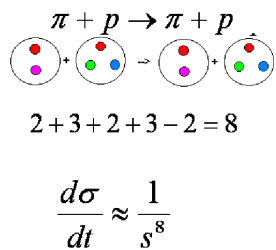

$$\pi + p \to \pi + p$$

$$2 + 3 + 2 + 3 - 2 = 8$$

$$\frac{d\sigma}{dt} \approx \frac{1}{s^8}$$

Fig. 2 Quark arithmetic for differential cross section of pion-nucleon scattering

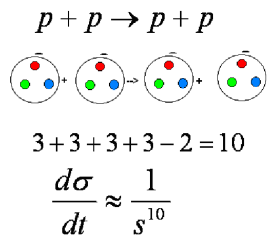

$$p + p \to p + p$$

$$3 + 3 + 3 + 3 - 2 = 10$$

$$\frac{d\sigma}{dt} \approx \frac{1}{s^{10}}$$

Fig. 3 Quark arithmetic for differential cross section of elastic proton-proton scattering

$$\gamma + d \to n + p$$

$$n_d = 6, \quad n_\gamma = 1$$

$$1 + 6 + 3 + 3 - 2 = 11$$

$$\frac{d\sigma}{dt} \approx \frac{1}{s^{11}}$$

Fig. 4 Photodesintegration of deuteron according to the Quark Counting Rules.

Thousand experiments proved the validity of Quark Counting Rules predictions. The great phenomenological success of dimensional Quark Counting Rules indicates that matter at short distances really exhibits quark structure. According to *inSPIRE* the paper [1] has been cited in nearly thousand theoretical and experimental works and co-cited more than ten thousand times [5], see also [6]-[8]. Recently, Polchinsky and Strassler derived the QCR in the framework of string theory [9], which has generated renewed interest in this topic. Brodsky pointed out, that phenomenological success of QCR, implies that Quantum Chromodynamics is a strongly coupled conformal theory at moderate energy, ignoring asymptotic logarithmic corrections.

3. PART II. SPIN/MASS SCALING RELATIONS FOR COSMIC OBJECTS

The most important characteristic of cosmic bodies are *mass m* and spin *J* (proper angular momentum). The origin of the angular momentum of cosmic objects and its relationship with mass of the object are definitely one of the most significant problems of physics and astrophysics.
It is interesting to note that Mons. Georges Lemaitre (1894-1966), former president of the Pontifical Academy of Sciences, and after him Edmund Whittaker have considered the Rotating Primeaval Atom as source of rotational motions in the Universe.
E.T. Whittaker in his essay 'Spin in the Universe', published in Yearbook of Royal Society, Edinburgh (1945) wrote: "Rotation is a universal phenomenon; the earth and all other members of the Solar System rotate on their axes, the satellites revolve round the planets, the planets revolve round the Sun, and the Sun himself is a member of Galaxy or Milky Way system which revolves in a very remarkable way. How did all this rotary motions come into being? What secures their permanence or brings about their modification? And what part do they play in the system of the world? "
Mass and spin are most fundamental quantum numbers of elementary particles too. In theoretical physics mass and spin are considered as two independent invariants of Poincare symmetry, i.e. Casimir invariants of ten parametric Poincare group. G. Chew and S. Frautschi where the first who has shown that between spin and mass of an object may be a fundamental connection and put forward celebrated Chew-Frautschi conjecture for hadrons. One of the urgent needs of theoretical physics and astrophysics today is obtaining greater insight into the problem of origin of spin /mass scaling relations for celestial bodies - planets, stars, galaxies and their systems. What is the ultimate source of spin and angular momentum in the Universe?

In elementary particle physics after the work of G. Chew and S. Frautschi it has become clear that spin $J$ and mass $m$ of hadrons are not independent but are inherently connected by simple relation

$$J(m^2) = J(0) + J'(0)m^2$$

where spin $J$ is measured in the units of Planck's constant $\hbar$ and slope $J'(0)$ has value $J'(0) \approx 1/(GeV/c^2)^2 \approx 1/m_p^2$.

Neglecting $J(0)$ at large $m$ this relation can be essentially rewritten in simpler form

$$J = \hbar \left( \frac{m}{m_p} \right)^2$$

Dimensional analysis and scaling considerations allowed extending this string-like (one-dimensional) formula to multi-dimensional case

$$J = \hbar \left( \frac{m}{m_p} \right)^{1+\frac{1}{n}} \quad n = 1, 2, 3$$

Below the numerical data must be considered in SI units.

$$\text{string (hadron)} \quad n=1: \quad J = \hbar \left( \frac{m}{m_p} \right)^2 = 3.769 \times 10^{19} \, m^2$$

$$\text{disk (galaxy)} \quad n=2: \quad J = \hbar \left( \frac{m}{m_p} \right)^{\frac{3}{2}} = 1.542 \times 10^6 \, m^{\frac{3}{2}}$$

$$\text{ball (star)} \quad n=3: \quad J = \hbar \left( \frac{m}{m_p} \right)^{\frac{4}{3}} = 53.11 \, m^{\frac{4}{3}}$$

Fundamental constants has value:

$m_p = 1.673 \times 10^{-27} \, kg = 0.938 \, Gev/c^2$ — proton mass

$\hbar = 1.055 \times 10^{-34} \, J \, s$ — Planck constant

$G = 6.673 \times 10^{-11} \, N \, m^2/kg^2$ — gravitational constant

$c = 2.998 \times 10^8 \, m/s$ — speed of light

$\dfrac{\hbar c}{G m_p^2} = 1.616 \times 10^{38}$ — dimensionless combination,

Gravitational or Kerr angular momentum $J_{Kerr} = Gm^2/c$ is a maximal angular momentum of rotating black hole with mass *m*. Here *c* is speed of light and *G* gravitational constant. Using Planck mass $m_{Pl} = \sqrt{\hbar c/G}$, it is possible to rewrite Kerr angular momentum in *string-like* form

$$J_{Kerr} = \frac{Gm^2}{c} = \hbar \left(\frac{m}{m_{Planck}}\right)^2 = 2.226 \times 10^{-19} m^2$$

$$\text{where } m_{Planck} = \sqrt{\frac{\hbar c}{G}} = m_p \sqrt{\frac{\hbar c}{Gm_p^2}} \text{ is a Planck mass}$$

This formula formally resembles relation for hadronic string like object after substitution of proton mass with Planck's one $m_p \to m_{Planck}$. There's a big difference in the slopes of trajectories, which is obvious from identity

$$J = \hbar \left(\frac{m}{m_p}\right)^2 \equiv \frac{\hbar c}{Gm_p^2} \frac{Gm^2}{c} = \frac{\hbar c}{Gm_p^2} J_{Kerr}$$

The dimensionless combination $\hbar c/Gm_p^2 = 1.7 \times 10^{38}$ exposes the great difference of slopes amongst hadronic and gravitational strings.

Equating Kerr momentum with spin relation for stars and galaxies gives two simple equations, from which coordinates of intersection points can be deduced [11]:

Coordinates of Chandrasekhar point: $\quad m_{star} = m_p \left(\frac{\hbar c}{Gm_p^2}\right)^{\frac{3}{2}} \quad\quad J_{star} = \hbar \left(\frac{\hbar c}{Gm_p^2}\right)^2$

Coordinates of Eddington point: $\quad m_{Universe} = m_p \left(\frac{\hbar c}{Gm_p^2}\right)^2 \quad\quad J_{Universe} = \hbar \left(\frac{\hbar c}{Gm_p^2}\right)^3$

The Chandrasekhar and Eddington limiting mass relations are considered as a brilliant achievement of theoretical physics. The corresponding expressions for limiting spins were found by present author [11]. Chandrasekhar's name was immortalized in connection with relation for $m_{star}$. One can state that *gravitation+quantum mechanics+hadron physics = Nobel Prize*

Let us consider double logarithmic plane. Any curve $y = c x^n$ in x,y plane is possible to recast as straight line in X,Y plane, where X= log x, Y=log y:

$$y = c x^n$$
$$\log y = \log c + n \log x$$
$$Y = C + n X$$

In these log-log plane theoretical spim/massrelations takes form:

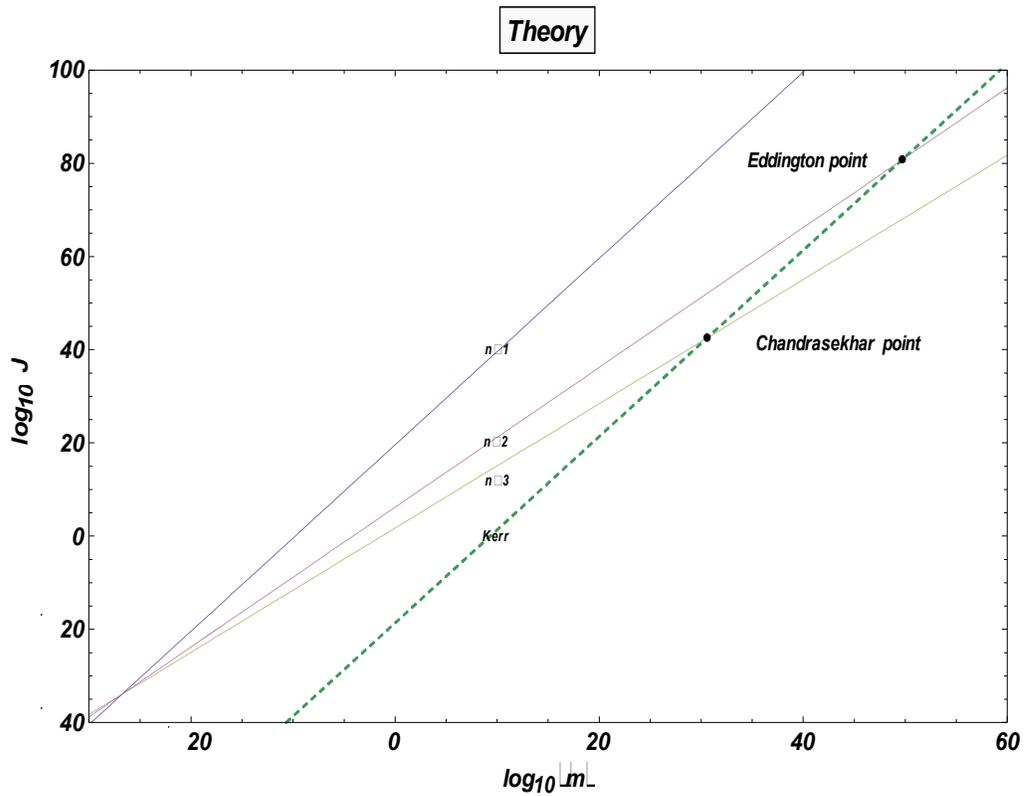

Fig.5 Theoretical construction on $log_{10}(J)$-$log_{10}(m)$ plane

Presently, the situation with observational data is very controversial. The pre dark matter data from [11] are represented in Fig. 6. Andromeda mass (including dark matter) is approximately $1.23 \times 10^{12}$ $m_\odot$ (or trillion solar masses). Probably Milky Way is more massive than M31. Milky Way mass (including dark matter) is approximately $1.9 \times 10^{12}$ $m_\odot$ According to our approach the predicted angular momentum of Andromeda is $5.79 \times 10^{69}$ $J\,s$ and of Milky Way is $1.11 \times 10^{70}$ $J\,s$.

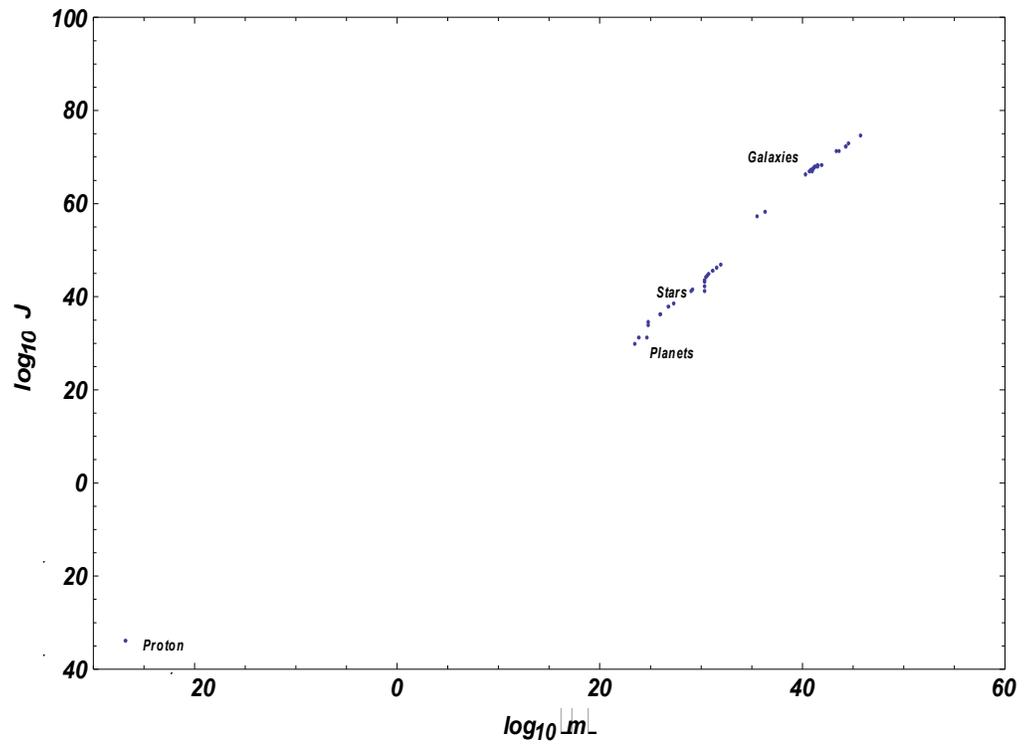
Fig. 7 Observational data: planets, stars, galaxies and proton

Together theoretical and observational data represents generalized Chew-Frautschi plot, where proton, laying on the intersection of three lines n=1,2,3, is a representative of hadronic world. The hadronic Regge trajectory is parallel to Kerr trajectory. Intersection of

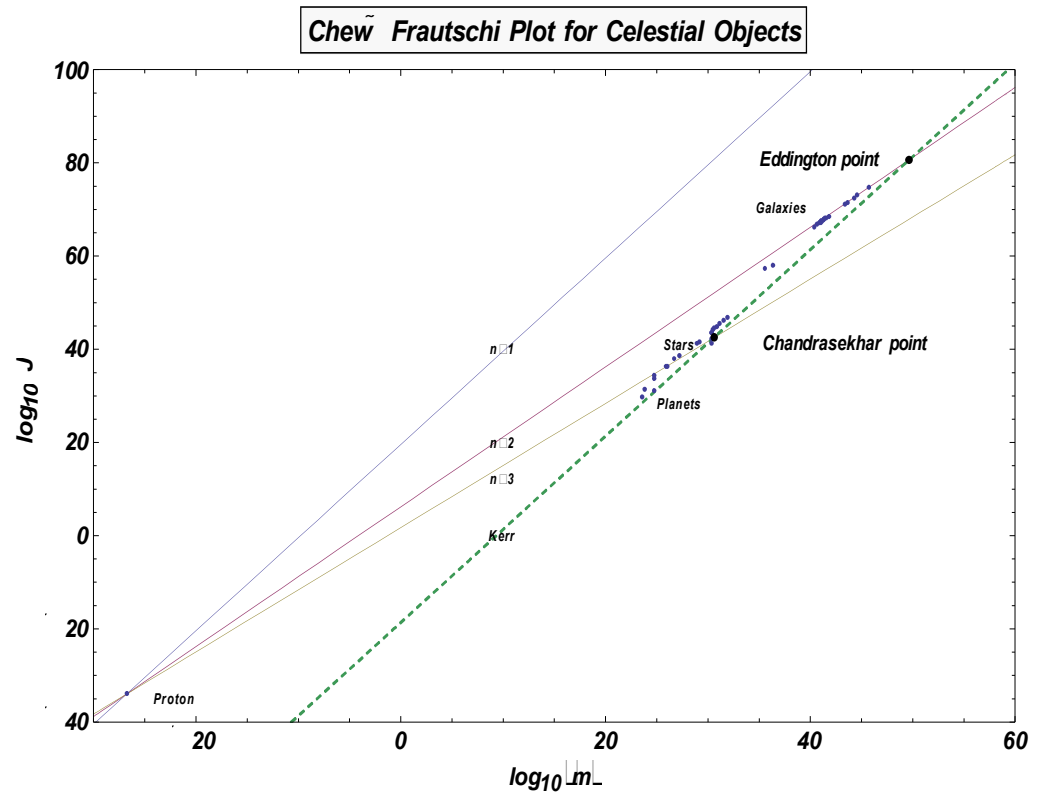

Fig. 8 Theoretical construction + Observational data

We have presented a new, *quantum-mechanical* model for the origin of the angular momentum of celestial bodies. Unlike to the previous classical attempts, our approach gives surprisingly accurate numerical predictions of the angular momentum for all spinning astrophysical objects. This occurs in a first time in the history of astronomy.

The creation of spin (angular momentum) is impossible through any applications of a classical field. Artificial invention (postulation) of torque fields such as shear is a unique way to create spin classically.

Another interesting result from this advance is merely philosophical and witness about the unity and simplicity of Nature in micro and macro scales. Understanding of this cannot be achieved by focusing narrowly on the *classical* side of subject. Instead it is necessary integrated, interdisciplinary, open-minded of quantum-*mechanical* vision of the problem of origin of rotation in astrophysics